\documentclass[prl,amsmath,amssymb,twocolumn,superscriptaddress,showpacs]{revtex4-1}

\usepackage{amsmath,amssymb}
\usepackage[usenames]{color}
\usepackage{amssymb}
\usepackage{grffile}
\usepackage[pdftex]{graphicx}
\usepackage{amsmath, amstext, amssymb, amsfonts, amsxtra}
\usepackage{textcomp}
\usepackage{xspace}
\usepackage{bbm}
\usepackage{bm}

\newcommand{\be}{\begin{equation}}
\newcommand{\ee}{\end{equation}}

\newcommand{\xiamen}{Department of Physics, Key Laboratory of Low Dimensional
Condensed Matter Physics (Department of Education of Fujian Province), and
Jiujiang Research Institute, Xiamen University, Xiamen 361005, Fujian, China}
\newcommand{\como}{Center for Nonlinear and Complex Systems, Dipartimento di
Scienza e Alta Tecnologia, Universit\`a degli Studi dell'Insubria, via
Valleggio 11, 22100 Como, Italy}
\newcommand{\infn}{Istituto Nazionale di Fisica Nucleare, Sezione di Milano,
via Celoria 16, 20133 Milano, Italy}
\newcommand{\brazil}{International Institute of Physics, Federal University
of Rio Grande do Norte, Campus Universit\'ario - Lagoa Nova, CP. 1613, Natal,
Rio Grande Do Norte 59078-970, Brazil}
\newcommand{\NEST}{NEST, Istituto Nanoscienze-CNR, I-56126 Pisa, Italy}

\begin{document}

\title{Onsager symmetry for systems with broken time-reversal symmetry}
\author{Rongxiang Luo}
\affiliation{\xiamen}
\author{Giuliano Benenti}
\affiliation{\como}
\affiliation{\infn}
\affiliation{\NEST}
\author{Giulio Casati}
\affiliation{\como}
\affiliation{\brazil}
\author{Jiao Wang}
\affiliation{\xiamen}

\begin{abstract}
We provide numerical evidence that the Onsager symmetry
remains valid for systems subject to a spatially dependent magnetic field,
in spite of the broken time-reversal symmetry.
In addition, for the simplest case in which the field strength
varies only in one direction, we analytically derive the result.
For the generic case, a qualitative explanation is provided.
\end{abstract}

%\pacs{05.70.Ln}
%05.70.Ln: Nonequilibrium and irreversible thermodynamics

\maketitle

{\it Introduction.--}
Onsager reciprocal relations \cite{onsager},
or the fourth law of thermodynamics, are a
cornerstone in nonequilibrium statistical physics.
These relations reflect on the macroscopic level
the time reversal symmetry of the microscopic dynamics.
That is, the equations of motion are invariant under the combined
reversal of time $t$ and momenta ${\bm p}$:
${\cal T}({\bm r},{\bm p},t)\equiv ({\bm r},-{\bm p},-t)$
(${\bm r}$ being the coordinates).
On a macroscopic level, this symmetry has striking consequences
on the phenomenological transport coefficients
\cite{callen,degrootmazur}.
Given a system brought out of equilibrium by the thermodynamic
forces $\mathcal{F}_k$,
the coresponding fluxes $J_k$ are such that in the linear
coupled transport equations $J_j=\sum_k L_{jk} \mathcal{F}_k$
the kinetic coefficient $L_{jk}$ obey the Onsager symmetry
$L_{jk}=L_{kj}$. For instance, in thermoelectricity \cite{Benenti2017}
$\mathcal{F}_e$ and $\mathcal{F}_h$ are the electrochemical and
temperature gradient,
$J_e$ and $J_h$ the charge and heat flow, and the Onsager symmetry
implies $L_{he}=L_{eh}$, or equivalently $\Pi=T S$, where
$\Pi=L_{he}/L_{ee}$ is the Peltier coefficient and
$S=L_{eh}/TL_{ee}$ is the Seebeck coefficient (or thermopower), $T$ being the
temperature. That is, as a consequence of time-reversal symmetry,
the Seebeck and Peltier effect can be treated on equal footing
and their interdependency is revealed.

On the other hand, the time-reversal symmetry ${\cal T}$ can be broken,
for instance by an applied magnetic field since the Lorentz force couples
coordinates and momenta. In this case, the laws of physics remain
unchanged under time reversal, provided that simultaneously
the magnetic field ${\bm B}$ is replaced by $-{\bm B}$:
${\cal T}_{\bm B}({\bm r},{\bm p},t,{\bm B})\equiv
({\bm r},-{\bm p},-t,-{\bm B})$.
This leads to the Onsager-Casimir relations \cite{onsager,casimir}
for the kinetic
coefficients: $L_{jk}({\bm B})=L_{kj}(-{\bm B})$. In the
illustrative example of thermoelectricity,
$\Pi({\bm B})=T S(-{\bm B}$), while in principle one could
have $\Pi({\bm B})\ne T S({\bm B})$ [that is,
$L_{eh}({\bm B})\ne L_{he}({\bm B})$], thus violating the
Onsager symmetry.
Equivalently, the Onsager-Casimir relations do not
impose the symmetry of the Seebeck coefficient (or of the Peltier coefficient)
under the exchange ${\bm B}\to -{\bm B}$, i.e.,
we could have $S({\bm B})\ne S(-{\bm B})$.

However, for the particular case of noninteracting systems
connected to two reservoirs, the relation $\Pi({\bm B})= T S({\bm B})$
is a consequence of the symmetry properties of the scattering matrix
\cite{datta}.
Moreover, for interacting systems subject to a \emph{constant} magnetic
field, it has been recently shown, both in classical
\cite{rondoni2014} and in quantum mechanics \cite{rondoni2017},
that the Onsager relations are still valid.
Then the relevant question arises: Under what general conditions
the Onsager relations remain valid?
For concreteness, is it possible to find
a nonuniform magnetic field and an interacting system
such that $\Pi({\bm B})\ne TS({\bm B})$ \cite{3terminal}?

In this letter, we provide convincing numerical evidence that,
for classical particles moving
in two dimensions (say, the $xy$ plane), the Onsager symmetry persists
for a generic magnetic field ${\bm B} = B(x,y) \,{\bm k}$, where
${\bm k}$ is the versor of the $z$ axis.
An analytical proof of the symmetry is given
for the case ${\bm B} = B(x) \,{\bm k}$
(or $B$ varying along any direction in the $xy$ plane),
while qualitative arguments are presented
for the generic case $B(x,y)$.

{\it Theory.--}
We consider a system of $N$ interacting particles, governed by the
Hamiltonian
\begin{equation}
H=\sum_i^N \frac{[{\bm p}_i-q_i{\bm A}({\bm r}_i)]^2}{2 m_i}
+\frac{1}{2}\sum_{i\ne j} V(r_{ij}),
\label{eq:hamiltonian}
\end{equation}
where ${\bm r}_i$ and ${\bm p}_i$ are the conjugated coordinates
and momenta of particle $i$ (of mass $m_i$ and charge $q_i$),
$V(r_{ij})$ is the interaction potential between particles
$i$ and $j$, and ${\bm A}$ is the vector potential.

We start by assuming ${\bm B}=B(x)\,{\bm k}$ (similar considerations
would apply for ${\bm B}$ varying along any direction in the
$xy$ plane). In what follows, we show that for systems exposed
to magnetic fields of this kind the dynamics is invariant under
the transformation
\begin{equation}
\begin{array}{c}
{\displaystyle
{\cal M}(x,y,z,p^x,p^y,p^z,t,{\bm B})}
\\
{\displaystyle \equiv
(x,-y,z,-p^x,p^y,-p^z,-t,{\bm B}).}
\end{array}
\label{eq:Msymmetry}
\end{equation}
Using the Landau gauge, we write the vector potential as
${\bm A}= A(x)\,{\bm j}$, with ${\bm j}$ versor of the
$y$ axis and $A(x)=\int_{x_0}^x B(x') dx'$,
the choice of $x_0$ being irrelevant.
It can be easily checked that Hamiltonian (\ref{eq:hamiltonian}),
and the equations of motion
\begin{eqnarray}
\left\{
\begin{array}{l}
{\displaystyle
\dot{x}_i=\frac{p_i^x}{m_i},
}
\\
{\displaystyle
\dot{y}_i=\frac{1}{m_i}\,[p_i^y-q_i A(x_i)],
}
\\
{\displaystyle
\dot{z}_i=\frac{p_i^z}{m_i},
}
\\
{\displaystyle
\dot{p}_i^x=F_i^x+\frac{q_i}{m_i}\,[p_i^y-q_iA(x_i)]\,B(x_i),
}
\\
{\displaystyle
\dot{p}_i^y=F_i^y,
}
\\
{\displaystyle
\dot{p}_i^z=F_i^z.
}
\end{array}
\right.
\end{eqnarray}
are preserved by symmetry (\ref{eq:Msymmetry})
($F_i^\alpha=-\frac{\partial
\sum_{j\ne i} V(r_{ij})}{\partial\alpha}$
represents the force, deriving from
particle-particle interactions,
on particle $i$ in direction $\alpha$).
It is important to remark that $V(r_{ij})$ is invariant
under transformation (\ref{eq:Msymmetry}) because we assume
that the potential depends only on
the distance between the particles.

The above symmetry considerations do not directly apply if
${\bm B}=B(x,y)\,{\bm k}$. Indeed,  if we choose the vector potential
as ${\bm A}=A(x,y)\,{\bm j}$, where
$A(x,y)=\int_{x_0}^x B(x',y) dx'$, we can see that
transformation (\ref{eq:Msymmetry}) implies
$p_i^y-qA(x_i,y_i) \to p_i^y-qA(x_i,-y_i)$
and in general $A(x_i,-y_i)\ne A(x_i,y_i)$.
To investigate this general case, we will therefore
resort to numerical simulations.

\begin{figure}[!t]
\vskip 0.5cm
\includegraphics[width=8.0cm]{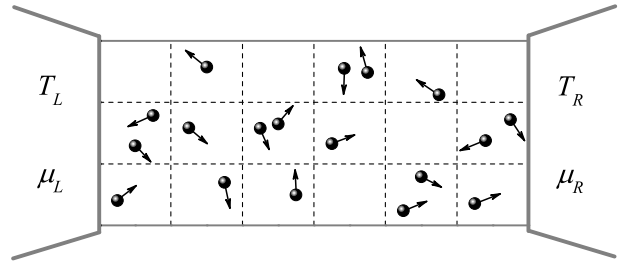}
\vskip 0.3cm
\caption{Schematic drawing of the 2D gas of interacting particles,
described by the multi-particle collision dynamics.
The cells of dashed-line boundaries represent the partition of
space used to model collisions. A magnetic field, transverse
to the plane of motion, is applied to the system
which is coupled to left and right electrochemical reservoirs.}
\label{fig:2DMPC}
\end{figure}

{\it Numerics.--}
We consider a two-dimensional
(2D) gas of interacting particles, of equal mass $m$ and charge $q$
(we set $m=q=1$). The particles are in a rectangular box of
length $L$ (along the $x$ coordinate) and width $W$
(along the $y$ coordinate), see Fig.~\ref{fig:2DMPC} for
a schematic plot.
The system is subject to a magnetic field
$B(x,y)$ directed along the $z$ axis.
The dynamics is described by the multi-particle collision (MPC)
dynamics~\cite{kapral}. The MPC simplifies the numerical simulation of
interacting particles by coarse graining the time and space
at which interactions occur.
By MPC, the system evolves in discrete time steps, consisting of
non-interacting propagation during a time $\tau$ followed by instantaneous
collision events. During the
propagation, each particle evolves under the Lorentz force
determined by the magnetic field.
For the collisions, the system is partitioned into identical square cells
of side $a$, then the velocities
of all particles found in the same cell are rotated with respect to
their center of mass velocity ${\bm V}_{CM}$
by two angles, $\alpha$ or $-\alpha$, randomly chosen
with equal probability. The
velocity of a particle in the cell is thus updated from ${\bm v}_i$ to
${\bm V}_{CM}+{\cal R}^{\pm \alpha}({\bm v}_i-{\bm V}_{CM})$, where
${\cal R}^{\theta}$ is the 2D rotation operator of angle $\theta$.
The collisions preserve the total energy and total momentum of
the gas of particles.

%%%%%%%%%%%%%%%%%
%%%%%%%%%%%%%%%%%

The system is placed in contact with two electrochemical reservoirs
at $x=0$ and $x=L$, through openings of the same size as the width $W$
of the box \cite{periodic}.
The left and right reservoirs are modeled as ideal gases and are characterized
by temperature $T_\gamma$ and electrochemical potential $\mu_\gamma$
($\gamma=L,R$).
We use a stochastic model of the reservoirs \cite{carlos2001,carlos2003}:
whenever a particle of the system crosses the boundary which separates
the system from the left or right reservoir, it is removed. On the other hand,
particles are injected into the system from the boundaries, with rates
and energy distribution determined by temperature and
electrochemical potential (see, \emph{e.g.}, \cite{Benenti2017}).
Thermoelectric transport was discussed with this method
\cite{Casati2008,Saito2010,Benenti2013,stark,Chen2015},
also for the MPC model \cite{Benenti2014,Luo2018}.

\begin{figure}[!t]
\includegraphics[width=8.0cm]{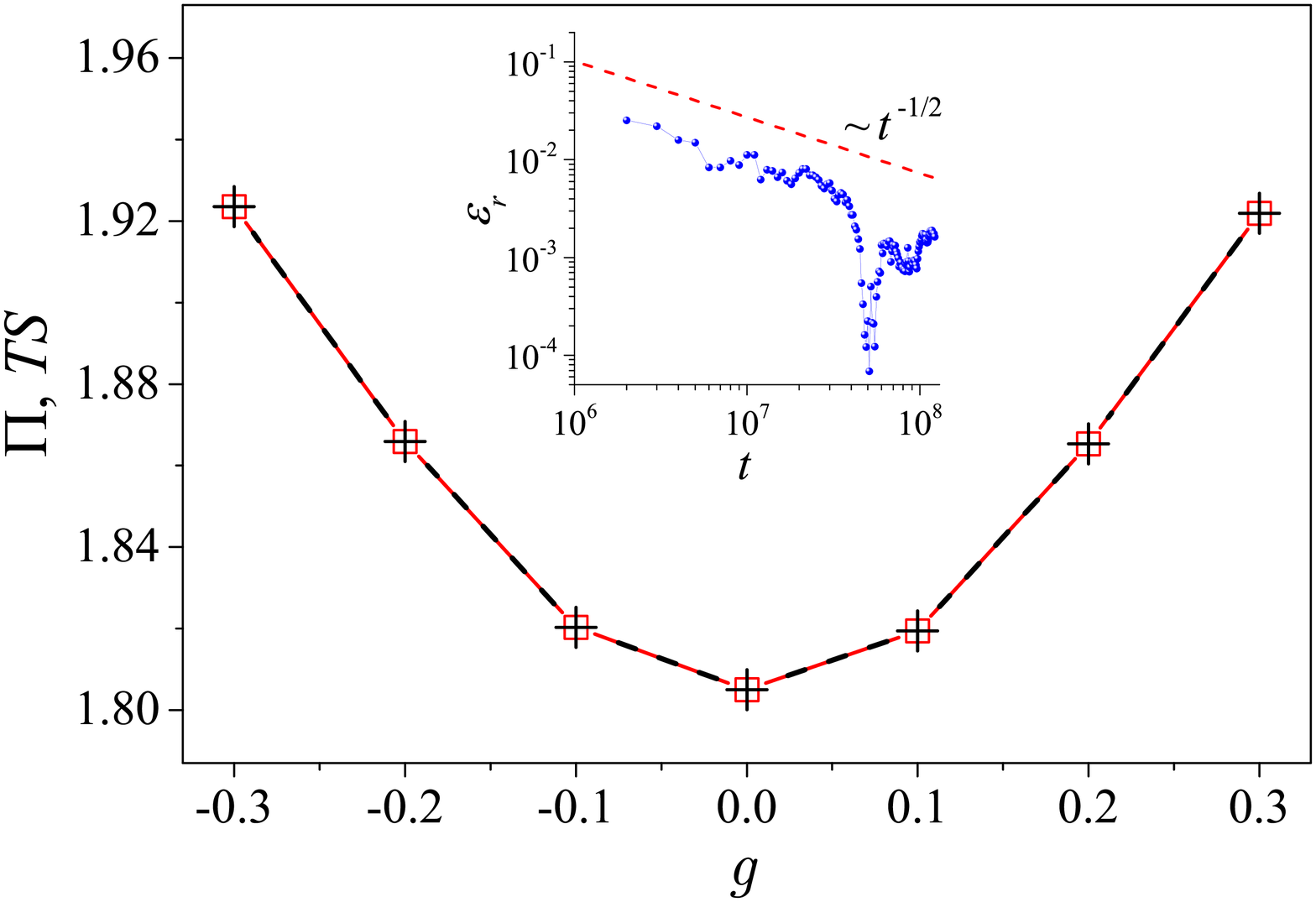}
\caption{Peltier coefficient $\Pi$ (red open squares) and
thermopower $S$ times temperature $T$ (black pluses)
as a function of $g$, for a field $B(x)=gx$.
Parameter values for the MPC simulations:
lenght $L=10$, width $W=2$, side of the square cells
$a=0.1$, time between collisions $\tau=0.25$, rotation angle
$\alpha=\pi/2$, temperature $T=1$, and particle density $\rho=22$ 
(the electrochemical potential is set to be $\mu=0$ for $\rho=22$ 
and $T=1$); the temperature and the electrochemical potential of 
the left (right) reservoir are, respectively, $T_L=T+\Delta T/2$ 
($T_R=T-\Delta T/2$) and $\mu_L=\mu+\Delta\mu/2$
($\mu_R=\mu-\Delta\mu/2$) with $\Delta T=0.05$ and $\Delta\mu=0.05$.
Inset: relative error $\varepsilon_r=|\Pi(g)-TS(g)|/\Pi(g)$ for $g=0.3$
versus integration time $t$ for $g=0.3$.}
\label{fig:Bx}
\end{figure}

We first consider the case $B(x)=g x$.
As expected from the above theory, the numerical results of
Fig.~\ref{fig:Bx} show that the Onsager symmetry $\Pi(g)=TS(g)$
is fulfilled for any value of $g$ (together with the
Onsager-Casimir relation $\Pi(g)=TS(-g)$, this implies
that the thermopower
$S(g)$ and the Peltier coefficient $\Pi(g)$ are even functions).
In the inset, we show the relative error
$\epsilon_r\equiv |\Pi(g)-TS(g)|/\Pi(g)$
for $g=0.3$. We can see that $\epsilon_r$,
due to the finite integration time $t$ in numerical
simulations, decreases $\propto 1/\sqrt{t}$, as expected for
statistical errors, and is smaller that $0.3\%$ for $t=1.2\times 10^8$.

%%%%%%%%%%%%%%%%%
%%%%%%%%%%%%%%%%%

\begin{figure}[!t]
\includegraphics[width=8.0cm]{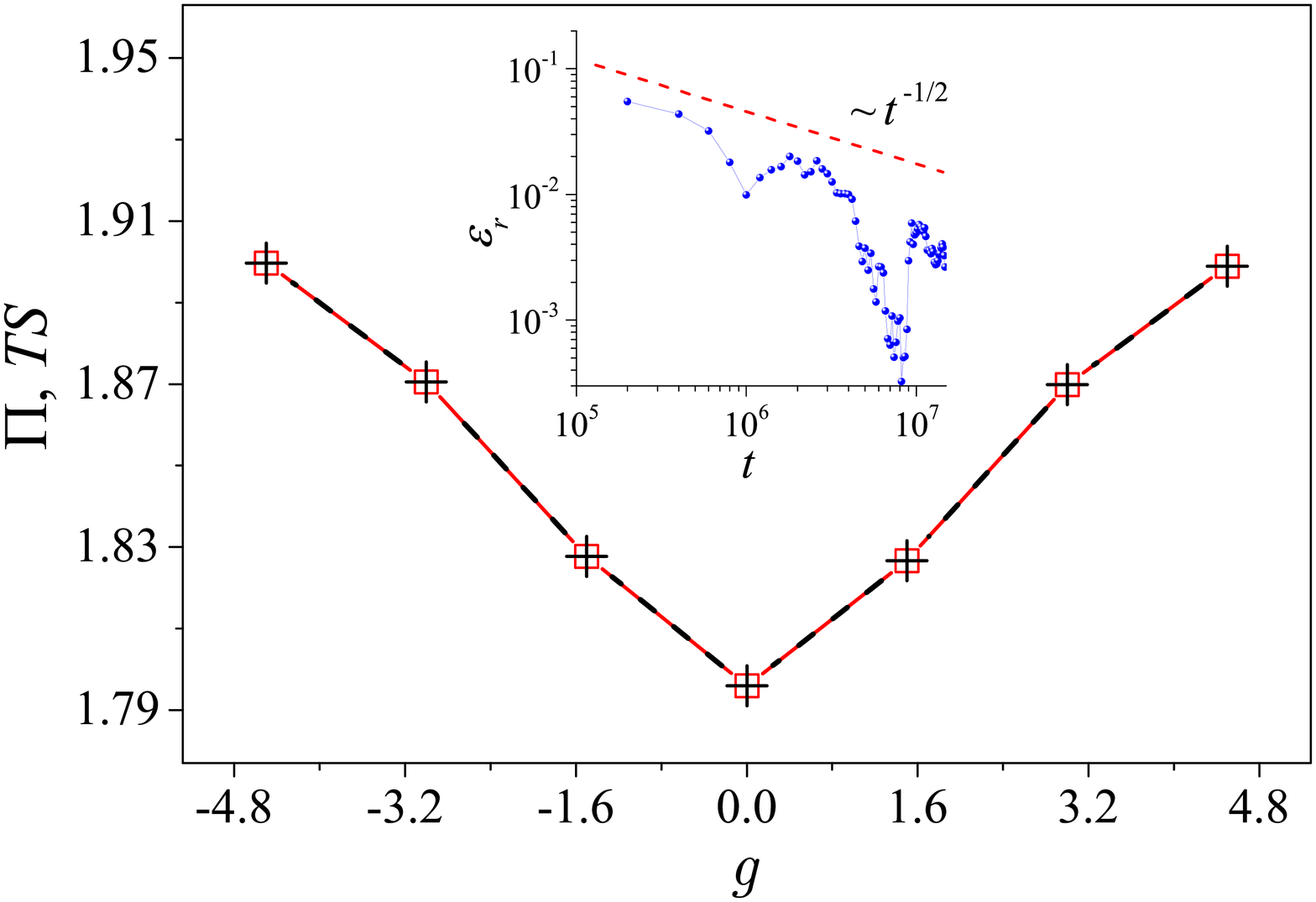}
\caption{Same as in Fig.\ref{fig:Bx}, but for the magnetic
field $B(x,y)=g\sin[\pi x/(2L)]\sin[\pi y/(2W)]$.
Parameter values are the same as for Fig.\ref{fig:Bx}.
Inset: relative error $\epsilon_r$ vs. integration time $t$
for $g=3$.}
\label{fig:Bxy}
\end{figure}

We then consider the generic case and numerically investigate
several functions $B(x,y)$,
without finding any statistically
significant violation of the Onsager symmetry.
As an illustrative example, in Fig.~\ref{fig:Bxy} we show results for
$B(x,y)=g\sin[\pi x/(2L)]\sin[\pi y/(2W)]$.
Similarly to the case of Fig.~\ref{fig:Bx}, the Onsager symmetry
is fulfilled, with the relative error $\epsilon_r\propto 1/\sqrt{t}$
(see the inset, where we show as an example $g=3$, for wich
$\epsilon_r$ is smaller that $0.5\%$ for $t=1.4\times 10^7$.

{\it Discussion and conclusions.--}
We have analytically shown that, for systems
in a magnetic field of strength varying along one direction,
there exists a symmetry such that the
equations of motion are invariant under time reversal without reversing the
magnetic field. As a consequence of such symmetry of the microscopic
dynamics, the Onsager reciprocal relations for the phenomenological
transport coefficients remain valid. On the other hand, extensive numerical
simulations carried out on two-dimensional systems suggest that the
symmetry persists for a generic $B(x,y)$ magnetic field.
This result can be qualitatively understood from the following argument \cite{rondoni}.
The field $B(x,y)$ can be approximated by a finite number $n$ of step
functions (in the $y$ direction), $B(x,y)\approx B(x,y_k)$
for step $k$ ($k=1,...,n$).
For each step the magnetic field is constant in the $y$ direction,
and therefore the above symmetry applies. On the other hand,
discontinuities of the field between steps would induce sudden changes of
velocity but not affect the symmetry.
The results of this paper could be extended to quantum mechanics,
with the proper counterpart of map ${\cal M}$ of Eq.~(\ref{eq:Msymmetry})
discussed in Ref.~\cite{rondoni2017}.
The question remains, for three-dimensional motion,
if the Onsager symmetry applies also
for a magnetic field
with both strength and direction depending on position.
We conjecture a positive answer on the basis of the following argument.
One could divide the system into small
volumes $dV_\alpha$ and for each volume
approximate the magnetic field with a constant vector.
Building a local Cartesian tern $(x_\alpha,y_\alpha,z_\alpha)$
for each $dV_\alpha$, with $z_\alpha$ pointing in the field direction,
symmetry (\ref{eq:Msymmetry}) applies locally.
For $dV_\alpha\to 0$,
%out of this local transformation for any trajectory
we expect
%, for ``regular'' enough magnetic fields,
to obtain a time-reversal trajectory without reversing
the magnetic field \cite{unpublished}.
%Furthermore, the effect on the Onsager symmetry of other time-reversal
%breaking effects like the Coriolis force should also be considered.

The results of this paper have consequences on the thermodynamic
bounds imposed on the efficiency of coupled transport.
The Onsager symmetry is a severe thermodynamic constraint
to the efficiency of (thermoelectric) energy conversion, and for
that reason it was suggested that a magnetic field, breaking that
symmetry, might be useful to enhance the performance of a thermoelectric
device \cite{Benenti2011}. The results of the present paper
exclude such possibility for two-terminal devices \cite{coriolis},
not only for noninteracting systems \cite{datta} but in
general transport models with strong particle-particle interactions.

%%%%%%%%%%%%%%%%%
%%%%%%%%%%%%%%%%%

{\it Acknowledgments:} We thank Keiji Saito for bringing to our attention
Ref.~\cite{rondoni2014} and Lamberto Rondoni for useful discussions.
We acknowledge support by NSFC (Grants No. 11535011
and No. 11335006) and by the INFN through the project QUANTUM.

%%%%%%%%%%%%%%%%%%%%%%%%%%%%%%%%%%%%%%%%%%%%
%%%%%%%%%%%%%%%%%%%%%%%%%%%%%%%%%%%%%%%%%%%%
%bibliography


\begin{thebibliography}{100}
\bibitem{periodic}
In the $y$ direction, the particles are subject
to fixed (hard wall) boundary conditions. On the other hand,
it has been verified numerically that the symmetry discussed in
this paper remains valid for periodic boundary conditions.
\bibitem{unpublished}
Numerical investigations to check our conjecture
are under way.
\end{thebibliography}
\end{document}